\documentclass[usenatbib]{mn2e}

\usepackage{graphicx}
\usepackage{dcolumn}
\usepackage{bm}
\usepackage{epstopdf}

\newcommand{\myskip}[1]{}

\newcommand{\mbar}{{\overline m}}

\newcommand{\gb}{{\bar g}}

\newcommand{\BEQ}{\begin{eqnarray}}
\newcommand{\EEQ}{\end{eqnarray}}
\newcommand{\BEA}{\begin{eqnarray}}
\newcommand{\EEA}{\end{eqnarray}}
\newcommand{\nn}{\nonumber}
\newcommand{\Sigmab}{\overline\Sigma}

\newcommand{\cm}{{\rm cm}}
\newcommand{\km}{{\rm km}}

\newcommand{\s}{{\rm s}}

\newcommand{\kpc}{{\rm kpc}}

\newcommand{\eV}{{\rm eV}}

\newcommand{\CG}{{\it cg}}
\newcommand{\co}{{\it co}}
\newcommand{\cut}{{\it cg}}
\newcommand{\Gal}{{\it G}}

\begin{document}

{

\title[Neutrino dark matter in A1689?]{Are observations of the galaxy cluster Abell 1689 consistent
with a neutrino dark matter scenario?}

\author[T.M. Nieuwenhuizen and A. Morandi]
{Theo M. Nieuwenhuizen$^{1}$\thanks{E-mail: t.m.nieuwenhuizen@uva.nl} and
Andrea Morandi$^{2}$\thanks{E-mail: andrea@wise.tau.ac.il} \\
$^{1}$Institute for Theoretical Physics, University of Amsterdam, Science Park 904, P.O. Box 94485,  1090 GL  Amsterdam, The Netherlands \\
$^{2}$ Department of Physics, Purdue University, 525 Northwestern Av.,
West Lafayette, Indiana 47907 - United States}

\maketitle

\begin{abstract}
Recent weak and strong lensing data of the galaxy cluster A1689 are modelled by dark fermions that are quantum degenerate within some core.
The gas density,  deduced from X-ray observations up to 1 Mpc and obeying a cored power law, is taken as input, while the galaxy mass density is modelled.
An additional dark matter tail may arise from cold or warm dark matter, axions or non-degenerate neutrinos.
The fit yields that the fermions are degenerate within a 430 kpc radius.
The fermion mass is a few eV and the best case involves 3 active plus 3 sterile neutrinos of
equal mass, for which we deduce $1.51 \pm 0.04$ eV. The eV mass range will be tested in the KATRIN experiment.
\end{abstract}


\section{Introduction}

The Lambda cold dark matter ($\Lambda$CDM) paradigm has been embraced after it was realized
that the most natural dark matter candidate, the neutrino, seems incapable to explain the formation
of large scale structures like galaxies and galaxy clusters.  
$\Lambda$CDM has been successful in describing the cosmic microwave background and large
scale structures. It is adaptable to many other situations, though often not predictive.
However, despite decades and dozens of searches, the CDM particle has not been established (Aprile et al  2012),
also not at the Large Hadron Collider (Caron 2011),
so one may wish to keep an open eye at other scenarios.

It has been put forward by Gibson (1996) that in the early Universe the role of turbulence and viscosity in the protoplasma
and after the transition to neutral gas is more important than commonly assumed.
Gravitational hydrodynamics alone is capable to explain large scale structure formation
without CDM trigger, in a three-step top-down scenario, see also Nieuwenhuizen et al. (2009).
First, the plasma undergoes a viscous fragmentation at redshift $z=5100$, creating voids of  40 Mpc comoving size.
After the decoupling of photons the gas condenses in Jeans clumps
of 600,000 solar masses and they in their turn fragment in 200 billion micro brown dwarfs of Earth weight. 
This picture explains a wealth of observations (Schild 1996, Nieuwenhuizen et al 2009, 2012) 
and motivates to re-open our minds for the possibility of free streaming dark matter, like neutrino hot DM.

The violent relaxation mechanism explains the success of isothermal models  in gravitation (Lynden Bell 1967), 
so it is natural to consider isothermal dark matter fermions.  Cowsik \& McClelland (1973)  model the DM of the Coma cluster
as an isothermal sphere of neutrinos,  which yields a mass of $\simeq 2$ eV.
Treumann et al (2000)  consider 2 eV neutrinos next to CDM and X-ray gas, all isothermal, and derive density profiles for, e. g.,  the Coma cluster. 
Nieuwenhuizen (2009) (N09) applies an isothermal fermion model for a single type of dark matter, cold or not, 
in addition to the galaxies and the X-ray gas in galaxy clusters. 
A fit to lensing data of the Abell 1689 cluster works well and yields as best case the neutrino 
with mass 1.45 eV, below the empirical upperbound of 2.0 eV (Amsler et al 2008). 
Active neutrinos (lefthanded $\nu$, righthanded $\bar\nu$) have density $102$ cm$^{-3}$
 in each of the 3 families  (Weinberg 2008), so with 1.5 eV mass they constitute about 10\% of the critical mass density.
Some 20\% can arise if the sterile partners
(righthanded $\nu$, lefthanded $\bar\nu$) have also been created in the early Universe (N09).

As these challenging results from the physics literature have hardly been noticed in the astrophysics community, it is immanent to test them further.
Here we shall apply the isothermal model for dark fermions to new data for the gas in A1689 from X-ray observations and
recent data for strong lensing (SL) and weak lensing (WL). We also compare to the fit of an NFW profile (Navarro, Frenk and White 1998).

\section{Description of the data}

For the strong lensing analysis we refer to Limousin et al (2007) who inferred the 2D mass distribution of A1689 by Hubble 
Advanced Camera for Surveys (ACS)  observations
and ground based spectroscopy. They implement a parametric analysis via the {\it Lenstool} package (Kneib et al 1996), and determine two large-scale dark matter
clumps, one associated with the center of the cluster and the other with a north-eastern substructure. The north-eastern sector of SL and X-ray
data is masked out in order to avoid the contribution from this secondary substructure. The 2D mass distribution is rebinned into
circular annuli, in order to measure the projected mass profile $\Sigma(r)$ and the covariance matrix  $C$ among all the measurements of $\Sigma(r)$.

For the X-ray analysis we refer to Morandi et al (2010),  who analyze 2 Chandra X-ray observations with a total exposure time of 150 ks. The vignetting-corrected
brightness image is extracted from the events 2 files in the energy range ($0.5$--$5.0$ keV). The gas density profile is recovered in a non-parametric
way by rebinning the surface brightness into circular annuli and via spherical deprojection (Morandi et al 2007). 
In order to infer the observed temperature profile, they implement the spectral analysis by extracting the source
 spectra from circular annuli around the centroid of the surface brightness. Once they assumed an absorbed MEKAL model, the spectral fit is performed
in the energy range 0.6--7 keV by fixing the redshift at $z_{A1689}=0.183$, and the photoelectric absorption at the galactic value. Background spectra, 
which are free from source emissions, are extracted from regions of the same exposure.
 
 The deprojected temperature is determined through spherical deprojection of the projected temperature profile inferred in the
spectral analysis (Morandi et al 2007).

We take WL data out to 3 Mpc from Fig. 16 of Umetsu and Broadhurst (2008), to be denoted as UB.
These authors derive a projected 2D mass map of A1689 by combining lens magnification
with distortion of red background galaxies in deep Subaru images. To derive the critical  surface density

\BEQ \Sigma_c=\frac{c^2}{4\pi G}\frac{D_s}{ D_dD_{ds}}
\EEQ
with $D_s$ the distance to the source galaxies, $D_d$ to the deflecting cluster and $D_{ds}$  their mutual distance.
UB employ  $\langle{D_{ds}}/{D_s}\rangle=0.693\pm0.02$, yielding $\Sigma_c=0.790\pm 0.023\,{{\rm gr}}\,{{\rm cm}^{-2}}$
for $h_{70}=1$, $\Omega_M=0.25$, $\Omega_\Lambda=0.75$ with $ z_s=0.68$
for the average redshift of the lensed galaxies in the WL analysis.

\section{Degenerate fermions}

We  investigate the role of thermal  fermions with mass $m$, degeneracy $\gb$, chemical potential $m\mu$ at
temperature $T_\nu=m\sigma_\nu^2$, located in a gravitational potential $\varphi(r)$

\BEQ \rho_\nu(r)\!=\!\int\frac{{\rm d}^3p}{(2\pi\hbar)^3}\,\frac{\gb m}{\exp\{[p^2/2m+m\varphi(r)-m\mu]/T_\nu \}+1}.
\EEQ
It is equal to ${\gb m}{\lambda_T^{-3}}\,{\cal L}i_{3/2}^-\left[(\mu-\varphi(r))/{\sigma_\nu^2}\right]$,
with thermal length $\lambda_T=\hbar\sqrt{2\pi}/m\sigma_\nu$ and  a polylogarithmic function ${\cal L}i^-_\gamma(x)=\sum_{k=1}^\infty k^{-\gamma}(-1)^{k-1}e^{kx}$
$\to 4x^{3/2}/3\sqrt{\pi}$ ($x\to\infty$).

To test the idea we assume spherical symmetry. 
 The density of free electrons is obtained by us in the interval $10$ kpc $<r<$ 1 Mpc
 and presented in Fig. 1. It fits well to a cored power law, better than to a $\beta$-profile. We thus employ

\BEQ \label{nefit}
n_e(r)=\frac{n_{e}^{0}(R_g^0)^a}{(r+R_g^0)^a}, \hspace{3mm}   a=2.38 \pm 0.05,
\EEQ
with $n_{e}^{0}=0.073 \pm 0.008$ cm$^{-3}$ and $R_g^0=124 \pm 7$ kpc.
For the 56 data points the fit is good, $\chi^2/\nu=43.9/53=0.827$.

\begin{figure}\label{nedata}
\includegraphics[scale=0.9]  {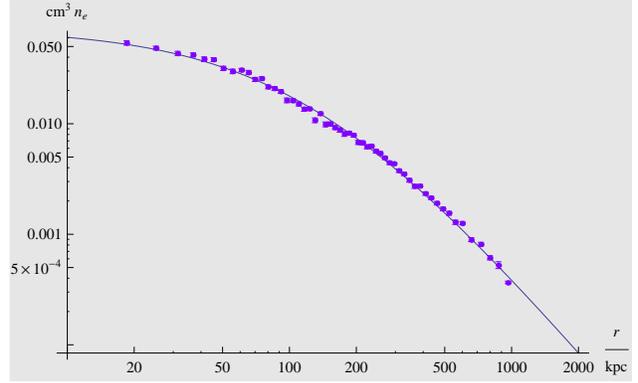}
\caption{The 56 data points with error bars for the electron density $n_e$ as function of $r$.
The fit is the cored power law (\ref{nefit}).}
\end{figure}

Fixing the H-He ratio at the cosmic value and the metallicity $Z=0.3$ retrieved by employing the solar abundance ratios from \cite{Grevesse1998},
we have an average molecular mass $\mbar=0.60\,m_N$, with $m_N$ the nucleon mass,  and a gas
particle density $n_g=1.852\,n_e$,  so that

\BEQ \label{ne-rhog-rel} 
\rho_g=\mbar n_g =\frac{\rho_{g}^{0}(R_{g}^{0})^a}{ (r+R_{g}^{0})^a}, \qquad \rho_{g0} = 1.11\,m_N \,n_{e}^{0}.
\EEQ
If this would hold beyond the observed range, the gas mass would diverge as $r^{3-a}$. To have only a logarithmical divergence,
we model the gas tail  by a Burkert (1995) profile,

\newcommand{\gas}{{\it gas}}
\newcommand{\dm}{{\it dm}}

\BEQ\label{Burkert}
\rho_g(r>r_1)=\frac{\rho_g^1 (R_g^1)^3}{(r+R_g^1)(r^2+(R_g^1)^2)}, \quad r_1=1\textrm{ Mpc}.
\EEQ
In the central case of Eq. (\ref{nefit}):  $\rho_{g}^{1}=0.0557\,\rho_{g}^{0}$, $R_{g}^{1}=587$ kpc.
The galaxy mass density is dominated by the central galaxy (cg). We model it as  in Limousin et al (2007), viz.

\BEQ
\rho_\Gal(r)
= \frac{\rho_\CG ^0 R_\co^2R_\CG^2}{(r^2+R_\co^2)(r^2+R_\CG^2)},
\quad 
\rho_\CG^0=\frac{\sigma_\CG^2(R_\co+R_\cut)}{2\pi GR_\co^2 R_\cut}.
\EEQ

\myskip{
Our main theme is to investigate the role of thermal fermions with mass $m$, degeneracy $\gb$ at chemical potential $m\mu$.
The approach appears to have $T=0$ as best fit.
The Fermi momentum $p_F(r)=m\sqrt{2(\mu-\varphi(r))}$ sets the mass density as

\BEQ \rho_\nu(r)\!=\!\bar gm\int_{p<p_F}\frac{{\rm d}^3p}{(2\pi\hbar)^3}=\frac{\sqrt{2}}{3\pi^2}\, \frac{\bar g m^4}{\hbar^3}\left(\mu-\varphi(r)\right)^{3/2}.
\EEQ}

Finally, we allow a dark matter component $\rho_\dm(r)$ of the form (\ref{Burkert}) with parameters $\rho_\dm^0$ and $R_\dm$.
It is present at all $r$, but mainly relevant in the outskirts.
It could stem from cold dark or warm dark matter, axions or from non-degenerate neutrinos.
The Poisson equation then reads

\BEQ
\varphi''(r)+\frac{2}{r}\varphi'(r)=4\pi G\rho(r), \quad \rho=\rho_G+\rho_g+\rho_\nu+\rho_\dm.
\EEQ
For the solution it is advantageous to split off the contributions from $\rho_G$, $\rho_g$ and $\rho_\dm$, that can be treated analytically. 

{} From strong lensing one determines the $2D$ mass density $\Sigma(r)=\int_{-\infty}^\infty{\rm d} z\rho\left(\sqrt{r^2+z{}^2}\,\right)$. Eq. (5) implies that 

\BEQ \label{SigmaFromPhip}
\Sigma(r)=
\frac{1}{2 \pi G} \int_0^\infty{\rm d} s\frac{ \cosh 2s}{\sinh^2s}
\left[\varphi'(r\cosh s)-\frac{\varphi'(r)}{\cosh^2s}\right].
\EEQ
Averaging it over a disk yields $\Sigmab(r)=2r^{-2}\int_0^r{\rm d} r'\,r'\Sigma(r')=M_{2D}(r)/\pi r^2$, which is equal to  
(Nieuwenhuizen 2009) 

\BEQ \Sigmab(r)=
\frac{1}{\pi G} \int_0^\infty{\rm d} s\, \varphi'(r\cosh s).
\EEQ

\myskip{
A reasonable fit of $\Sigma$ data can be found if we take the $T_g$ data and employ the modified
hydrostatic equilibrium condition $(n_gT)'=-5.5 n_g\bar m \varphi'$.
}

In weak lensing one determines the shear, which is related to $\Sigmab$ and $\Sigma$ as  
(see Eq. 7 of Umetsu et al.  2011)

\BEQ\label{gt=}
g_t(r)=\frac{\Sigmab(r)-\Sigma(r)}{\Sigma_c-\Sigma(r)}.
\EEQ
We evaluate $\chi^2_{WL}=\sum_i [g_t(r_i)-g_{t,i}^{\rm obs}]^2\sigma_{g,i}^{-2}$, in which we include also the data with $r_i<400$ kpc, where both $\Sigma$'s 
 in Eq. (\ref{gt=}) are important. We adopt the average of upper and lower errors, $\sigma_{g,i}=\frac{1}{2}(\sigma_{g,i}^{(+)}+\sigma_{g,i}^{(-)})$.
For the strong lensing it holds that  $\chi_{SL}^2=\sum_{i,j=1}^{12}[\Sigma(r_i)-\Sigma^{\rm obs}_i] ( C'{}^{-1})_{ij}[\Sigma(r_j)-\Sigma^{\rm obs}_j]$.
While $C$ is the covariance matrix,  we employ the modified covariance matrix $C'=\sigma_{SL}^2I+C$. 
In our parametric approach, $C$ has eigenvalues between $0.014$ and $6.5\cdot10^{-9}$ gr$^2$cm$^{-4}$, the low ones being unphysical. 
We therefore use Tikhonov regularization, adding a diagonal term $\sigma_{SL}^2$ accounting for further scatter and cutting off the
small eigenvalues. We adopt the typical value from the diagonal elements,  $\sigma_{SL}=(\sum_{i=1}^{12}C_{ii}/12)^{1/2}=0.0362$ gr cm$^{-2}$.
Next we minimize $\chi^2=\chi^2_{SL}+\chi^2_{WL}$ with the FindMinimum routine of Mathematica, speeded up by first searching at $\sigma_\nu=0$.
The best fit parameters with linear regression error bars are

\newcommand{\cg}{{\it cg}}
\BEQ&&
\hspace{-0mm}
\frac{\bar g}{12} \frac{(mc^2)^4}{\eV^4}=5.17\pm0.58, \quad
\mu= (33.6 \pm 1.4) {\sigma_{500}^2} , \\ &&
M_\cg=(3.62\pm 0.30)10^{13}M_\odot, \quad
R_\co= 7.9 \pm 1.1\, \kpc, \nn\\&&
\rho_{dm}^0=(0.0186 \pm 0.0044)\frac{m_N}{\cm^3},  \nn
\EEQ
where $\sigma_{500}=500\,\km\,\s^{-1}$. 
These values are taken at the optimal parameters  $\sigma_\nu=0.96\sigma_{500}$, $R_\dm =564$ kpc, $R_\cg=129.5$ kpc.
If we allow them to vary in the fit, they attain large error bars, $1.9\sigma_{500}$, 560 kpc, and 260 kpc, respectively, and induce larger errors in (11).
However, since the neutrino temperature and velocity dispersion are ill constrained, we may as well consider to work  in the $T_\nu=\sigma_\nu=0$ limit, or in our optimal case.
Next, the scale of the DM tail $R_\dm=0.96R_g^1$ basically overlaps with the scale $R_g^1$ on which the gas decays,  as is expected, so there is no reason to consider other values.
Finally, our  galaxy cut radius $R_\cg\approx R_g^0$ coincides with the $128.5\pm37.0$ kpc of Limousin et al (2007), so their analysis constrains it better.
Still, their core radius $R_\co=5.2\pm1$ kpc is somewhat smaller than ours, while 
their central velocity dispersion  $\sigma_\cg$ $=$ $370.5$ $\pm10.3$ km s$^{-1}$ is well below our our $\sigma_\cg=\sqrt{GM_\cg/\pi R_\cg}=619  \pm 25\, \km \, \s^{-1}$,
and hence their central mass  $M_\cg=8.8\,10^{12}M_\odot$ lies below ours, see (11).

The SL fit in Fig. 2 matches the data well, with $\chi^2_{SL} = 5.88$ only, despite the small error bars.  
The fast decline beyond 400 kpc is absent in the stacking of  the SL+WL data with those of 3 other heavy clusters (Umetsu et al 2011).
The WL data in Fig. 3 are well matched, also in the regime  $r<400$ kpc, where the $\Sigma$ in the denominator of (\ref{gt=}) is relevant,
and for $r>1$ Mpc with modelled gas density. The fit has
$\chi^2_{WL} = 6.77$ only. For our 12 + 11 data points and 5 parameters we thus have  $\nu=18$ and $\chi^2/\nu=0.68$;
being below 1, the fit may even be ``too good''. In fig. 3 we also show the best fit without DM tail, it has $\chi^2/\nu=1.89$.

\begin{figure}
\label{figSLUBdata}
\includegraphics[scale=0.9]{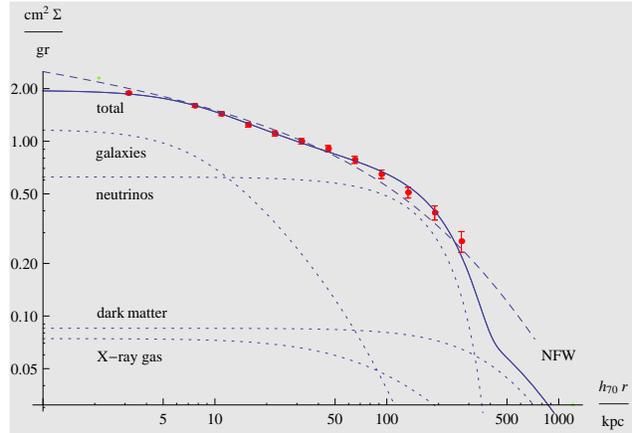}

\caption{Line-of-sight mass density $\Sigma$ as function of $r$.
Full line: best fit to our model. Dotted lines: separate contributions from galaxies, neutrinos, dark matter tail and X-ray gas.
Dashed line: best fit to an NFW profile.
Data from Limousin et al (2007); error bars from diagonal elements of the modified covariance matrix. }
\end{figure}

\begin{figure}
\label{figWLUBdataa}
 \includegraphics[scale=0.9]{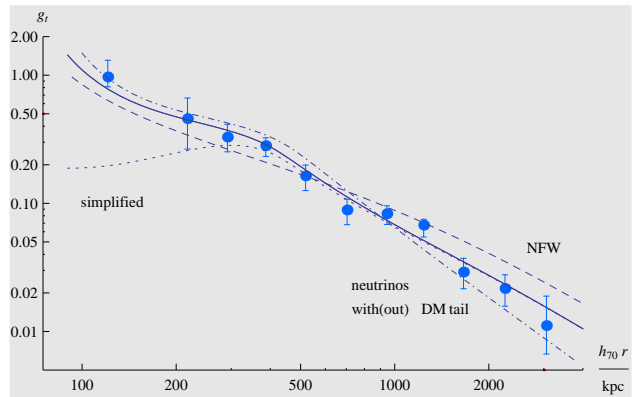}
\caption{
Weak lensing shear $g_t$ as function of $r$.
Full line: best fit to the neutrino model with a dark matter tail.
Dotted line:  Eq. (\ref{gt=}) without $\Sigma(r)$ in the denominator. 
Dashed-dotted line: best fit without  dark matter tail.
Dashed line: best fit to an NFW profile. Data taken from Umetsu and Broadhurst (2009).}
\end{figure}

The fermion mass emerges in the eV range, likely to be neutrinos. 
Considering 3 active + 3 sterile neutrinos of equal mass, they have $\bar g=12$ degrees of freedom and  a mass

\BEQ \label{mnu=}
m=1.51\pm 0.04\,\, {\rm eV},
\EEQ
which overlaps within $2\sigma$ with the $1.45\pm0.03$ eV of N09.

\newcommand{\vir}{{200}}
The average mass density is equal to 200 times the critical density at the cluster redshift for $r_\vir =1.98$ Mpc, 
where $M_\vir =(14.6\pm 1.5) 10^{14} M_\odot$, of which $28.3\pm5.5\%$ is in degenerate neutrinos, $53.5 \pm 7.3\%$  in dark matter, 
$15.8\pm1.6\%$ in gas and and $2.4\pm0.3\%$ in Galaxies. 
This can be compared with the estimate $M_\vir =14.1^{+6.3}_{-4.7}10^{14}M_\odot$ (Bardeau, 2005) and $(13\pm4)10^{14}M_\odot$ (Lemze et al, 2009). 
At 875 kpc the enclosed mass  $(8.3\pm 0.4)10^{14}M_\odot$  compares well with the Riemer-S\o rensen et al (2009) lensing value $(8.6\pm 3.0)10^{14}M_\odot$.
The resulting $18.2\pm1.9$\% of matter in baryons is near the cosmic fraction $  4.4/(4.4+21.4) = 17.4 \%$, as expected.
Leaving out the DM tail would imply a too large ($\sim26\%$) baryon fraction, and the larger $\chi^2/\nu=1.89$.

In Fig.  4 we plot the separate components of the mass density. The galaxy density crosses the neutrino density at  41 kpc. 
The ``quantum-to-classical" transition ${\cal L}i_{3/2}^-\left[(\mu-\varphi(r))/{\sigma_\nu^2}\right]=1$ occurs at
 $r_{qc}=431\,h_{70}^{-1}$ kpc; the neutrinos are quantum degenerate  for $r<r_{qc}$.
Due to exclusion principle, their density has no central cusp.

The gas temperature data, presented in fig. 5, can be viewed in the light of an hydrostatic equilibrium (HE).
With pressure $p_g=n_gT_g$, the HE condition $p_g'=-\rho_g\varphi'$ implies

\BEQ 
\label{Tg=}
k_BT_g(r)=\frac{\mbar}{\rho_g(r)}\int_r^\infty {\rm d}w\,\rho_g(w)\varphi'(w).
\EEQ
Because the empirical $T_g$ is fairly constant, HE demands that $\rho(r)\approx {\rm const}\times\rho_g(r)$.
While this is expected in the outskirts, see fig. 4,  it is problematic in the centre, where  the galaxies dominate, $\rho_G\gg\rho_g$. Hence the gas is likely not in HE.
Indeed, in Fig. 4 the order of magnitude is correct, but the observed values are exceeded up to a factor 2.5.

While lensing provides information on the instantaneous configuration,
the gas hydrodynamics may involve gas flowing in, which also expresses itself in the discrepancy between virial mass and X-ray mass. 
The latter has been diminished by dropping hydrostatic equilibrium (Molnar et al 2010)  or incorporating triaxiality (Morandi et al 2010).

After fitting the $T_g$ data to a cubic polynomial in  $\log r$, we can define a $\varphi_{\it HE}$ by 
$\varphi'_{{\it HE}}=-k_B(\rho_gT_g'+\rho_g'T_g)/\mbar \rho_g$ and from it a  $\Sigma_{{\it HE}}$ via (\ref{SigmaFromPhip}). 
At small $r$ it basically coincides with the data in Fig. \ref{figSLUBdata}, but as $r$ increases it becomes smaller, 
by a factor $2.5$ for the last 4 data points. Still, for $r\sim1$ Mpc $\Sigma_{\it HE}$ and $\Sigma$ approach each other, as expected.
Hence the gas is not in hydrostatic equilibrium below 1 Mpc, where Lemze et al (2011) find different velocities of DM and galaxies.

\begin{figure}
\label{rhoGalgasnu}
 \includegraphics[scale=0.9]{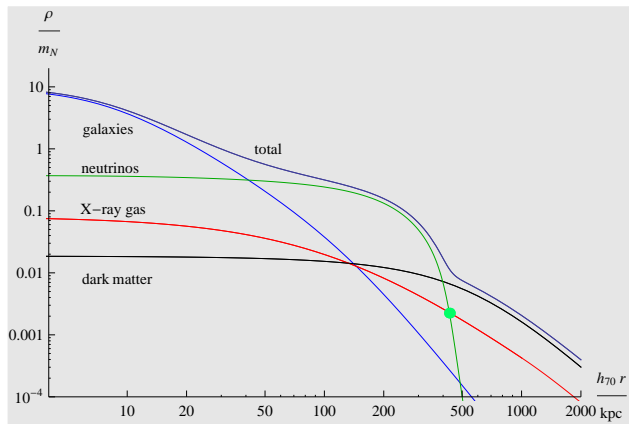}
\caption{The $3D$ mass density $\rho(r)$, decomposed in its contributions from galaxies,  neutrinos,  X-ray gas and dark matter.
The neutrinos are quantum degenerate for $r<r_{qc}=430$ kpc, indicated by the dot, where the gas density happens to coincide}
\end{figure}

 \begin{figure} \label{figTNFW}
 \includegraphics[scale=0.9]{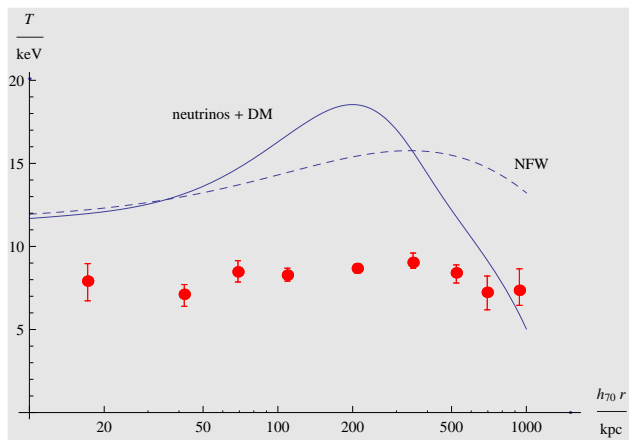} 
 \caption{X-ray temperature profile as derived from hydrostatic equilibrium within the neutrino model and with an NFW profile. Data from our gas analysis.} 
 \end{figure}

\section{NFW modelling of the data}

Cold dark matter can be modelled by the NFW profile, which involves the critical mass density $\rho_c=3H_0^2/8\pi G$, 

\BEQ
\label{NFWprof}
\rho_{\rm NFW}(r)=\frac{200c^3\rho_c\, (1+z_{A1689})^{3}}{3[\log(1+c)-c/(1+c)]}\,\frac{r_s^3}{r(r+r_s)^2}
\EEQ
The galaxy and gas densities should be added to it.
However, a competition between the central peaks of the NFW and of the galaxies yields bad fits, $\chi^2\sim700$, for reasonable values of $R_\co$, $R_\cg$ and $M_\cg$.
Hence we continue with describing {\it all} matter by NFW. $\Sigma$ and $\overline{\Sigma}$ thus follow from (\ref{NFWprof}) with $\varphi'=GM(r)/r^2$.
The fit to SL alone yields $\chi^2_{SL}=6.5$, $r_s=626$ kpc and $c_\vir =4.4$, while
the fit to WL alone yields $\chi^2_{WL}=6.4$, $r_s=179$ kpc and $c_\vir =11.0$, the latter in accord with UB.
These known discrepancies (see, e. g., UB; Lemze et al 2009) 
cast  a shadow on the combination of  SL + WL.
Indeed, this brings  a domination by the small  SL errors, viz.
$\chi^2_{SL}=19.8$, $\chi^2_{WL}=26.1$, $r_s=433$ kpc and $c_\vir=5.48$.
Its goodness-of-fit is only 0.0013, while its $\chi^2/\nu=45.8/19=2.18$ exceeds our previous $0.68$ by a factor 3.

The NFW temperature prediction is obtained by inserting $\varphi'=GM(r)/r^2$ in (\ref{Tg=}). The result is depicted in Fig. 5. 
It exceeds the measured temperatures even at 1000 kpc, so it is neither a reasonable fit, showing once more that the
HE problem is not solved within NFW modelling.

\section{Discussion}

This Letter presents new data for the electron density and temperature in the X-ray gas of A1689. Combined with
strong lensing data of Limousin et al (2007) and weak lensing data of Umetsu and Broadhurst (UB, 2008), 
this allows to test the isothermal fermion model of Nieuwenhuizen (2009), upon adding a DM tail.
Based on partly old data, it achieved to model the cluster dark matter as isothermal fermions.
The model could easily have been inadequate at the new level of information,
but it appears that the previous finding is mainly reproduced.
Both the strong and weak lensing data are well described.

On the other hand, the NFW profile of cold dark matter, which performs well for Sl and WL separately,
appears to perform less well when both data sets are combined. 
This contrasts to UB, who used older SL data.
Not only is $\chi^2/\nu$ about 3 times larger than for neutrinos, the quality of the NFW fit is also lower due to
systematic deviations for WL.

The reason why isothermal neutrinos work well may lie in the fact that their temperature-to-chemical-potential ratio is small,
 $T_\nu/\mu m=\sigma_\nu^2/\mu=0.027$, so that essentially one considers quantum degenerate $T_\nu=0$ neutrinos, a ``neutrino star",
 of 0.86 Mpc in diameter.

In the present work the gas density of A1689 is taken from observations up to 1 Mpc.
After general indications for this possibility (Kull et al. 1996), 
it was predicted that in A1689 the $\nu$DM is more localised in the centre than the gas (Nieuwenhuizen 2009). 
This partial segregation between DM and gas is reminiscent of the situation in the bullet cluster.
It is confirmed in the present approach, see Fig. 4, where  the neutrino core is more localised than the gas and DM.

Observations of galaxy clusters generally suggest them to contain fewer baryons (gas plus stars) 
than the cosmic fraction. This ``missing baryon'' puzzle is particularly relevant for massive clusters,
like A1689, as they are expected to represent the cosmic matter content (baryons and dark matter). 
It has recently been solved.
For a relatively large sample of groups and clusters of galaxies the gas fraction has been 
measured within $r_{500}$ and it is concluded that the ``missing'' baryons are present as X-ray gas 
located in the outskirts of the clusters (Rasheed et al 2011).
Our gas data have allowed an extrapolation which confirms this picture.

Other support for the neutrino picture may come from the ``cosmic train wreck" galaxy cluster Abell 520,
where a lot of DM is located in the centre with the X-ray gas, both separated from the galaxies that are out of the center
(Jee et al 2012, Clowe et al 2012). While this is hard to explain within $\Lambda$CDM,
cores made up of quantum degenerate matter that undergo a central collision may coalesce in the center.

For neutrinos all $\gb=12$ left- and righthanded states are available in the cluster.
Alternatively, this hints at 3 + 3 families of nearly equally massive  Majorana neutrinos.
This implies a mass $ m=1.51\pm0.04\,{\rm eV}$ and a cosmic DM fraction $\Omega_\nu=9.8\%$ for active neutrinos, and if the
sterile ones with this mass are also generated in the early Universe, $\Omega_\nu=19.5\%$.
This may be achieved by a Majorana mass matrix with meV entries, next to a
Dirac matrix with the $\sim1.5$ eV mass eigenvalues (Nieuwenhuizen 2009).
It will also produce neutrino-less double $\beta$-decay, with $m_{\nu\nu}\sim 1 $ meV.

In the early Universe neutrinos maintain  the quasi-relativistic Fermi-Dirac distribution $[\exp(pc/k_BT_\nu)+1]^{-1}$ with $T_\nu(z)=(4/11)^{1/3}T_\gamma(z)$,
until the neutrinos condense (``$\nu$c'') on the baryons in the galaxy cluster (Nieuwenhuizen 2009).
This happens when the ``Hubble force'' $H\langle p_\nu\rangle=3.151\, Hk_BT_\nu/c$ equals Newton's $Gm_\nu(M_\vir ^{\it Gal}+M_\vir ^{\it gas})/r_\vir ^2$.
The corresponding redshift is  $z_{\nu{\rm c}}\sim 7$, below the estimate of Nieuwenhuizen (2009). 
The neutrino mass contained at $z_{\nu{\rm c}}$ in a sphere of radius $r_{100}=2.65$ Mpc can be estimated from the cosmic DM density,
$(4\pi/3)[0.2 \rho_c(z_{\nu{\rm c}}+1)^3]r_{100}^3=11\, 10^{14} M_\odot$, near the $M_\vir ^\nu+M_\vir^{dm}=(4.1+7.8) $ $10^{14}M_\odot$ captured according to our model.
This supports the possibility that the DM tail consists of neutrinos as well.
 The  $T_\nu$ = 45 mK in A1689 lies below the would-be free streaming {\it momentum} temperature of 1.95 K
but, as it should,  above the related  0.96 mK  {\it kinetic} temperature 
(Nieuwenhuizen 2009). 

Gravitational hydrodynamics is a top-down scenario where cosmic voids and proto-clusters are formed early 
on in the plasma and it is consistent with neutrino hot dark matter (Nieuwenhuizen, Gibson \& Schild 2009).
Neutrinos would be localised near the centre of the Virgo supercluster to which the Local Group belongs, and hardly occur in the Galaxy.
Galactic dark matter would be baryonic, constituted by ``missing'' baryons, for which direct  and indirect indications exist (Nieuwenhuizen et al 2010, 2012). 

The mass of the electron antineutrino will be searched in the 2015 KATRIN
experiment from the present upper bound of 2 eV down to 0.2 eV. 
The uncertainty is $\Delta m_{\bar\nu_e}^2{} _{\rm sys} = 0.017\, \eV^2$ (Weinheimer 2009). 
If a mass near $1.5$ eV exists, this implies $\Delta m_{\bar\nu_e}{} _{\rm sys} = 6$ meV 
or  $\Delta m_{\bar\nu_e}{} _{\rm sys}/{m_{\bar\nu_e}} = 0.4\%$, well below our $2.5\%$ 
statistical inaccuracy.
The average neutrino mass is much larger than the oscillations, so they play no role:
the solar value $\Delta m^2_{32} = (2.43 \pm 0.13) \, 10^{-3} {\rm eV}^2$ implies $\Delta m_{32} /m_{\bar\nu_e}=0.048\%$. 

If in KATRIN a mass of  eV order is indeed observed, this will rule out the $\Lambda$CDM picture.
Its supposed exclusion of neutrinos as the dominant non-baryonic dark matter can then be traced back to
the assumption of CDM. Structure formation then has to start out nonlinearly and be top-down, 
welcoming very early galaxies and large scale structures like the axis of evil (Nieuwenhuizen et al 2010).

In conclusion, we model the strong and weak lensing of the galaxy cluster A1689.
The NFW profile performs moderately for the combined data sets.
A central core of quantum degenerate neutrinos combined with a dark matter tail performs well and confines the neutrino mass.
This dark matter tail may be composed of the usual suspects, cold or warm dark matter, 
axions, or ... non-degenerate neutrinos.
To explain the gas temperature remains an outstanding problem.

{\it Acknowledgments}:
We thank Marceau Limousin and H\aa kon Dahle for discussion and for assistance
with the data acquisition, and the unknown referee for careful remarks.
This work was supported in part by the US Department of Energy through
Grant DE-FG02-91ER40681 and Purdue University. 

\vspace{-0.6cm}
\newcommand{\aj} {AJ}
\newcommand{\apj} {ApJ}
\newcommand{\apjl} {ApJL}
\renewcommand{\bf}{}

\end{document}